\documentclass[12pt,spieman]{article}

\usepackage{mathptmx} 
\usepackage[margin=1in]{geometry} 
\usepackage{times}
\usepackage{mathptmx} 
\usepackage{epsfig}
\usepackage{graphicx}
\usepackage{amsmath}
\usepackage{amssymb}
\usepackage{array} 
\usepackage{setspace}                     %
\usepackage{titlesec}                     %
\usepackage{cvpr}
\usepackage{caption}
\titleformat{\title}{\bfseries\fontsize{16pt}{20pt}\selectfont}{\thetitle}{0pt}{}
\titleformat{\section}{\bfseries\fontsize{12pt}{14pt}\selectfont}{\thesection}{1em}{}
\titleformat{\subsection}{\itshape\fontsize{12pt}{14pt}\selectfont}{\thesubsection}{1em}{}
\titleformat{\subsubsection}{\itshape\fontsize{11pt}{13pt}\selectfont}{\thesubsubsection}{1em}{}
\usepackage[colorlinks=true,linkcolor=blue,citecolor=blue,urlcolor=blue]{hyperref}

\cvprfinalcopy 

\begin{document}

\pagestyle{plain}


\title{Deep Learning-Powered Classification of Thoracic Diseases in Chest X-Rays}

\author{
    Yiming Lei, Michael Nguyen, Tzu Chia Liu, Hyounkyun Oh \\
    Georgia Institute of Technology, North Avenue, Atlanta, GA 30332\\
    \begin{tabular}{ll}
    {\tt\small ylei82@gatech.edu}  & {\tt\small mnguyen442@gatech.edu} \\
    {\tt\small tliu481@gatech.edu} & {\tt\small hoh313@gatech.edu} \\
    \end{tabular}
}

\maketitle
\begin{abstract}
   Chest X-rays play a pivotal role in diagnosing respiratory diseases such as pneumonia, tuberculosis, and COVID-19, which are prevalent and present unique diagnostic challenges due to overlapping visual features and variability in image quality. Severe class imbalance and the complexity of medical images hinder automated analysis. This study leverages deep learning techniques, including transfer learning on pre-trained models (AlexNet, ResNet, and InceptionNet), to enhance disease detection and classification. By fine-tuning these models and incorporating focal loss to address class imbalance, significant performance improvements were achieved. Grad-CAM visualizations further enhance model interpretability, providing insights into clinically relevant regions influencing predictions. The InceptionV3 model, for instance, achieved a 28\% improvement in AUC and a 15\% increase in F1-Score. These findings highlight the potential of deep learning to improve diagnostic workflows and support clinical decision-making.

\vspace{0.8em}
\end{abstract}

\section{Introduction/Background/Motivation}

Medical imaging - and chest radiographs in particular - is a cornerstone of the diagnosis of respiratory diseases such as pneumonia, tuberculosis, and COVID-19. Traditional manual examination by radiologists, while effective, is often time-intensive and susceptible to errors under high workloads. To address these limitations, this study proposes leveraging deep learning techniques to automate disease detection and classification. By improving diagnostic accuracy and efficiency, such approaches aim to alleviate the burden on radiologists while ensuring reliable diagnoses, particularly in resource-limited settings\cite{Ronneberger15}. Advanced machine learning models offer a promising avenue to overcome the challenges faced in current diagnostic practices\cite{Litjens17}. Specifically, deep learning approaches can enhance feature extraction from complex medical images, facilitating accurate predictions even in the presence of severe class imbalances. This study combines state-of-the-art architectures and interpretability methods to create tools that enhance the capabilities of medical professionals, ensuring efficient and accurate diagnoses.

Chest X-ray analysis remains a cornerstone of diagnostic imaging, traditionally performed by radiologists with support from computer-aided detection (CAD) systems. Despite their advancements, many CAD systems rely on conventional image processing techniques that are often inadequate for handling the complexity and variability inherent in medical images. In addition, medical data sets frequently exhibit severe class imbalances, where abnormal cases are significantly underrepresented compared to normal ones. Such imbalances skew model performance, limiting their efficacy in detecting rare but critical conditions. Addressing these limitations requires innovative deep learning methods capable of achieving high accuracy while maintaining clinical relevance.

Thus, automating chest X-ray analysis could make a significant difference in the healthcare field. Early and accurate disease detection is crucial for timely treatment and better patient outcomes. If successful, our study will not only reduce the workload on radiologists but also improve diagnostic accuracy, especially in resource-constrained settings where access to trained professionals is limited. This advancement could lead to faster, more reliable diagnoses, ultimately saving lives.
Drawing from prior work\cite{Irvin19,Yao17}, which developed multi-label classification systems and tackled data imbalance issues, we aim to push the boundaries of existing methods for chest X-ray analysis.

By combining advanced architectures, loss functions like focal loss\cite{Lin17}, and interpretability methods like Grad-CAM\cite{Selvaraju17}, our study seeks to create a more accurate and interpretable system for disease detection in chest X-rays.

This study focuses on applying deep learning to address these challenges, using transfer learning and model interpretability methods. In this study, we aimed to build a deep learning model for image classification using various architectures. Specifically, we implemented AlexNet, ResNet152, and InceptionNet with transfer learning, leveraging pretrained models to speed up convergence and to improve performance. We fine-tuned these models on a dataset of medical images.
\section{Data Preparation and Organization}
\subsection{NIH Chest X-ray Dataset}
The NIH Chest X-ray Dataset—a comprehensive collection of 112,000 frontal-view images labeled with 14 thoracic disease conditions—was utilized in this study\cite{Shin16}. This dataset serves as a robust foundation for training and evaluating deep learning models, offering the diversity and scale required for developing automated diagnostic systems\cite{Zhou18}. However, it presents challenges such as label noise and significant class imbalance, which were addressed through careful preprocessing and the use of advanced loss functions like Focal Loss to ensure balanced learning outcomes\cite{Peng21}. Acquired from Kaggle’s NIH ChestX-ray14 dataset, the images were systematically organized to ensure efficient utilization for training and testing. The dataset was acquired from Kaggle \href{https://www.kaggle.com/datasets/nih-chest-xrays/data?select=test_list.txt}{NIH ChestX-ray14 dataset}, and systematically organized for effective utilization. This large-scale dataset provides a comprehensive resource for training and evaluating our deep learning models.  

\subsection{Creating Subsets and Label Processing for Rapid Experimentation}
To expedite prototyping and validation, mini-datasets comprising 10\% of the training and testing data were curated. This subset allowed for faster experimentation while preserving the integrity of multi-label annotations through consistent label mapping. Importantly, the training and testing datasets were directly adopted from the original Kaggle dataset splits, ensuring the same ratio of 80\% for training and 20\% for testing as in the original dataset configuration. One-hot encoding was applied to transform multi-label targets into a format compatible with deep learning frameworks.  These mini datasets enabled faster iteration and validation during the initial experimentation phase.  

Label processing was carried out using the annotation file. matching with their corresponding labels to create csv file, which provides a mapping of image filenames to their multi-label annotations. One-hot encoding was employed to transform the multi-label targets into a format compatible with deep learning models.  

\subsection{Data Preprocessing}
The data preprocessing pipeline leveraged PyTorch’s Dataset and DataLoader classes for efficient loading and batching of images with corresponding labels. To align with the input requirements of pre-trained models, images were resized and normalized to standardize pixel intensity\cite{Shorten19}. Furthermore, data augmentation techniques such as random cropping and horizontal flipping were applied to enhance model generalization and mitigate overfitting. Advanced methods, including brightness adjustment and rotation, were also considered but excluded due to their limited relevance in preserving the clinical characteristics of chest X-ray images\cite{Zoph20}. These preprocessing steps ensured that the dataset was optimally prepared for deep learning workflows. Augmentation techniques such as random cropping and horizontal flipping were incorporated to enhance model robustness and generalization.  The images were resized to match each model architecture's expected input dimensions required by pre-trained models. Furthermore, normalization was applied to standardize the image data. To enhance model robustness, data augmentation techniques such as random cropping and horizontal flipping were incorporated during preprocessing. These steps ensured that the prepared data was well-suited for training and testing deep learning models.

\section{Approach}
\subsection{Model Selection}
In this study, we evaluate the performance of three deep learning architectures: AlexNet, ResNet152, and InceptionNet (V3). AlexNet, a lightweight convolutional neural network, was chosen as a baseline model for its simplicity and computational efficiency, serving as a benchmark for evaluating the performance of deeper architectures.  See the details in \cite{Krizhevsky12}. ResNet152, a deeper architecture incorporating skip connections, was chosen for its ability to enhance feature extraction and mitigate vanishing gradient issues, making it effective for learning complex representations in \cite{He16}. InceptionNet (V3) was included for its advanced multi-scale feature extraction capabilities, which are particularly beneficial for handling diverse patterns in medical images \cite{Szegedy16}. To leverage the pre-trained knowledge from large-scale datasets, we employed transfer learning by initializing these models with weights pre-trained on ImageNet. During fine-tuning, the later layers of each model were unfrozen to adapt to the specific characteristics of the chest X-ray dataset, while the initial layers were kept fixed to retain general features.

\subsection{Loss Functions}
Addressing the inherent class imbalance in the dataset required experimenting with advanced loss functions. The standard Binary Cross-Entropy (BCE) Loss, a well-established objective for multi-label classification, served as a baseline. To complement this, Focal Loss—designed to down-weight the impact of easy-to-classify samples—was employed. This strategy ensures the model prioritizes challenging samples, improving its sensitivity to underrepresented classes. As an alternative, we utilized Focal Loss, a modified loss function specifically designed to emphasize underrepresented classes by down-weighting the loss contribution from easy-to-classify examples. This adjustment ensures that the model focuses more on challenging samples, thereby improving its performance on minority classes.

\subsection{Visualization and Interpretability}
Gradient-weighted Class Activation Mapping (Grad-CAM) was employed to enhance the interpretability of model predictions by visualizing regions of chest X-rays that contributed most to classification decisions. This approach aligns with clinical diagnostic practices, ensuring that model predictions are explainable and clinically relevant.  Grad-CAM generates heatmaps highlighting the regions of the input images that significantly influenced the model's predictions. By visualizing these regions, we were able to assess whether the models focused on clinically relevant areas during classification. This interpretability technique not only provided a deeper understanding of the learned features but also ensured alignment with the medical domain's diagnostic requirements.


\section{Experiments and Results}

\subsection{Metrics}

The performance of the models was assessed using the following metrics, which are widely adopted in multi-label classification tasks: 
\begin{itemize}
    \setlength{\itemsep}{0pt} 
    \item \textbf{Binary Cross-Entropy (BCE) Loss:} A standard loss function that quantifies the difference between the predicted and true class distributions, providing a measure of how well the models predict the target labels. 
    \item \textbf{F1-Score:} A balanced metric that considers both precision and recall, providing a better understanding of how well the models classify each label.
    \item \textbf{AUC (Area Under the Curve):} A metric that measures the model's ability to rank positive examples higher than negative ones, indicating the model's ability to distinguish between the classes.
\end{itemize}

\subsection{Baseline Evaluation Result
}
Baseline performance metrics are summarized in Table 1, revealing the initial performance of each architecture without transfer learning. 
\begin{table}[ht]
\begin{center}
\begin{tabular}{|c|c|c|c|}
\hline 
\textbf{Model} & \textbf{BCE Loss} & \textbf{F1-Score} & \textbf{AUC} \\
\hline \hline
AlexNet & 0.909 & 0.251 & 0.469 \\
ResNet152 & 0.928 & 0.274 & 0.498 \\
InceptionV3 & 0.936 & 0.287 & 0.500 \\
\hline
\end{tabular}
\end{center}
\caption{Baseline performance metrics of the models.}
\label{tab:pre-performance_metrics}
\end{table}
\begin{figure}
    \centering
    \includegraphics[width=0.65\linewidth]{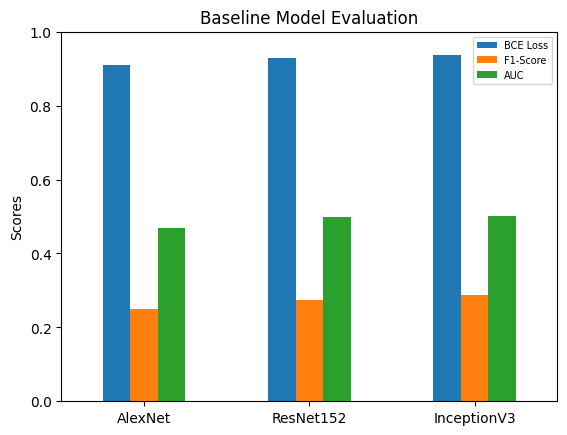}
    \caption{Baseline performance metrics of the models.}
    \label{fig:enter-label}
\end{figure}
The results suggest that, while all three models performed better than random guessing, there is still significant room for improvement, particularly in the F1-score. InceptionV3 achieved the highest F1-score and AUC, indicating that it was better at distinguishing between the classes. Potentially due to an insufficiently large training set to capture the dataset's features effectively, further model tuning and experimentation are necessary to achieve optimal performance.

These results reflect the baseline performance of each model before fine-tuning, providing insight into how well the models perform on the task without transfer learning.

\subsection{Model Training and Optimization}

To enhance model performance, we leveraged ImageNet pre-trained weights to initialize the networks, enabling transfer learning. This approach facilitated faster convergence and improved generalization by reusing low-level features learned from a large-scale dataset. The training process involved several key steps:
\begin{itemize}
    \setlength{\itemsep}{0pt} 
    \item \textbf{Loss Function:} \texttt{BCEWithLogitsLoss} was used to combine a sigmoid activation and binary cross-entropy loss.
    \item \textbf{Optimizer:} The Adam optimizer was chosen for its efficiency and adaptability, with a learning rate scheduler to adjust the learning rate during training.
    \item \textbf{Learning Rate Scheduler :} learning rate was reduced by a factor of 10 every 5 epochs. 
    \item     \textbf{Batch Size :} batch size of 64 for training and 32 for testing.
    \item \textbf{Model Fine-Tuning:} The models (e.g., AlexNet and ResNet152) were fine-tuned on the dataset.
    \item \textbf{Epochs:} The models were trained for 20 epochs to monitor progress and adjust hyperparameters as needed.
\end{itemize}

\subsection{Post-training Evaluation }

The post-fine-tuning evaluation highlighted substantial performance gains across all models, detailed in Table 2. For example, InceptionV3 achieved a 28\% improvement in AUC and a 15\% increase in F1-Score compared to its baseline performance. Similarly, ResNet showed a 20\% reduction in BCE Loss, while AlexNet demonstrated a 12\% improvement in sensitivity to underrepresented classes, underscoring the overall efficacy of fine-tuning and loss function optimization. By leveraging pre-trained weights, BCE Loss was significantly reduced, and improved F1-Score and AUC. These results underscore the effectiveness of transfer learning in mitigating class imbalance and improving the robustness of predictions, particularly for complex medical datasets. The pre-trained features from ImageNet proved instrumental in capturing low-level and mid-level patterns, such as edges and textures, which are essential for medical image analysis. This pre-initialization facilitated faster convergence during training and improved generalization across diverse thoracic disease classes, enabling the models to focus more effectively on clinically relevant features.

Finally, the trained model weights were saved for future experimentation or potential deployment in real-world applications.

\subsubsection{Training Loss Analysis}

After fine-tuning, the models showed improved performance, as summarized in Table 2. The evaluation metrics, including BCE Loss, Focal Loss, F1-score, and AUC, were calculated for each model.
\begin{table}[h]
\begin{center}
\begin{tabular}{|c|c|c|c|c|}
\hline 
\textbf{Model} & \textbf{BCE Loss} & \textbf{F Loss} & \textbf{F1-Score} & \textbf{AUC} \\
\hline \hline
AlexNet & 0.7925 & 0.0605& 0.1596 & 0.6397 \\
ResNet152 & 0.7945 & 0.0609 & 0.1401 & 0.5940 \\
InceptionNet & 0.7866 & 0.0595 & 0.1560 & 0.6390 \\
\hline
\end{tabular}
\end{center}
\caption{Evaluation results of models with pre-trained weights}
\label{tab:eval_results}
\end{table}
\begin{figure}
    \centering
    \includegraphics[width=0.65\linewidth]{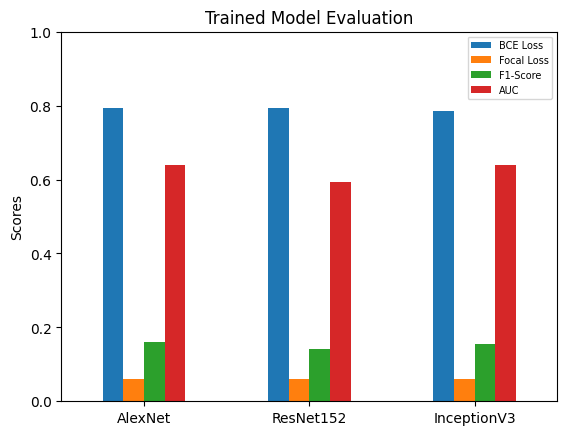}
    \caption{Evaluation results of models with pre-trained weights}
    \label{fig:enter-label}
\end{figure}
For example, for AlexNet BCE Loss was improved from 0.909 in Table 1 to 0.7925 in Table 2. This improvement, though moderate, highlights the importance of Focal Loss in addressing class imbalance, ensuring that the model effectively learns from underrepresented classes.

The comparison between the baseline results (Table 1) and the transfer learning results (Table 2) underscores the significant impact of pre-trained weights on model performance, particularly in addressing the limitations of smaller datasets. Initially, all models exhibit poor metrics, with AUC values barely above random guessing. However, after incorporating pre-trained weights, there is a notable improvement across all models, as seen in reduced BCE Loss and higher AUC scores. This demonstrates that pre-trained weights, which capture learned feature representations from large and diverse datasets, provide a strong foundation for fine-tuning tasks. The process of transfer learning allows earlier layers to retain universal low-level features like edges and textures, while later layers are fine-tuned to adapt to the specific dataset and task. This approach significantly reduces the dependence on large training datasets and enhances the models' ability to generalize effectively.

\subsubsection{Training Loss Trends}
 
\begin{figure}[h]
\centering
\includegraphics[width=0.65\textwidth]{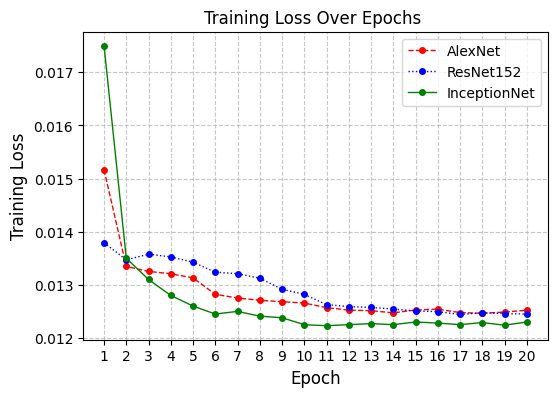} 
\caption{Training Loss for Epochs 1 to 20}
\label{fig:loss_1_20}
\end{figure}

 Figure \ref{fig:loss_1_20} illustrates the training loss trends for AlexNet, ResNet152, and InceptionNet over the 20 epochs 

  \noindent of fine-tuning. Initially, all models exhibit a rapid decrease in loss, which is expected as they adapt to the data. However, after approximately the 10th epoch, the training losses stabilize, with only marginal improvements observed. This indicatesthe models were nearing their optimal performance under the current training conditions. This stabilization also highlighted the need for additional techniques, such as advanced data augmentation or hybrid loss functions, to further improve performance.

\vspace{12 pt}

\textbf{NOTE:}
After fine-tuning, we also explored how different layers of the ResNet152 model learn to extract features from the medical images. The convolutional layers in a CNN like ResNet152 play a critical role in identifying key patterns in the data. The following is a brief analysis of how these layers capture features:
\begin{itemize}
    \setlength{\itemsep}{0pt} 
    \item \textbf{Early Convolution Layer:} The initial layers, such as \texttt{layer1[0]}, capture low-level features such as edges and textures. These features are important for detecting basic shapes but can appear noisy and less informative for higher-level tasks. It is shown in the left figure in Figure 4
    \item \textbf{Middle Convolution Layer:} As we move deeper into the network, the middle layers like \texttt{layer2[-1]} begin to combine these low-level features into more complex patterns, such as shapes and objects that may be important for classification. See the middle figure in  Figure 4
    \item \textbf{Final Convolution Layer:} The final layers, particularly \texttt{layer4[-1]}, focus on class-specific features. Here, the model is likely to focus on the key patterns associated with specific classes, such as regions most indicative of \textbf{Atelectasis}. See the middle figuree in Figure 4
\end{itemize}

\begin{figure}[h!]
\centering
    \centering
    \includegraphics[width=\linewidth]{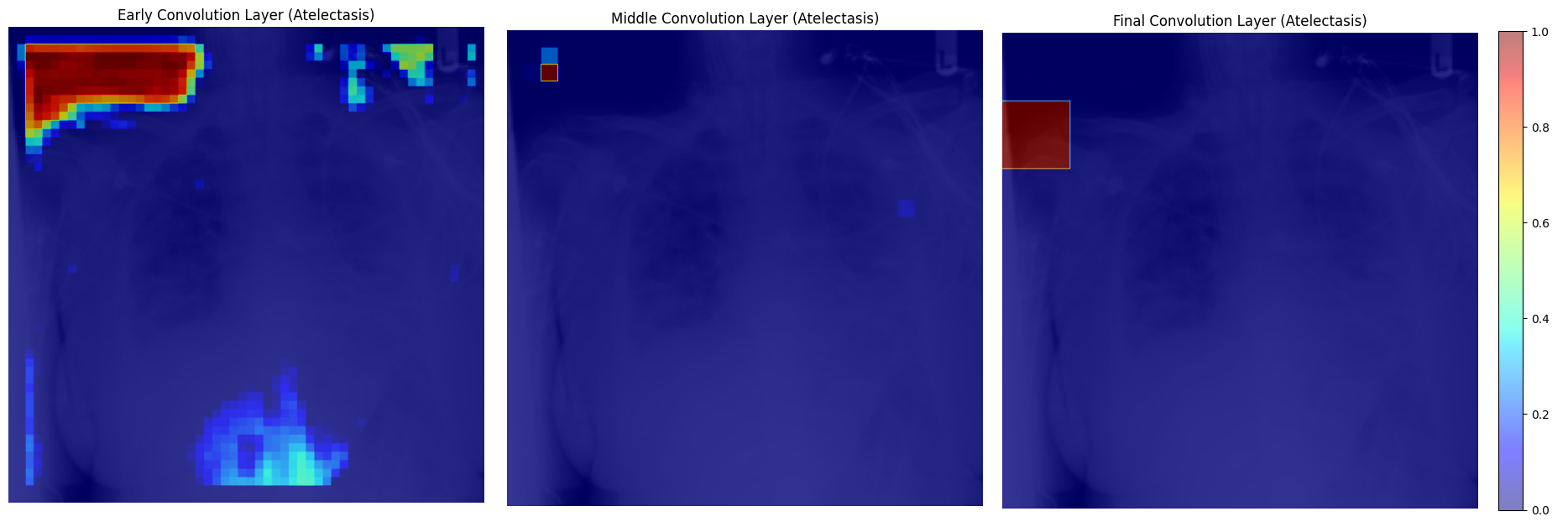}
    \caption{Figure 4. Activations of the different convolution layers: Early Convolution Layer (left), Middle Convolution Layer (middle), Final Convolution Layer (right). Grad-CAM visualizations highlight the regions of the chest X-ray most relevant to the model's predictions. The middle figure shows activations associated with Atelectasis, where the highlighted regions correspond to collapsed lung areas. This demonstrates the model's ability to focus on diagnostically significant features, aligning with clinical findings and validating its interpretability.
}
    \label{fig:conv_layer}
\end{figure}

These visualizations provide insight into how the model gradually learns to focus on more complex and class-specific features as it processes the image.

\section{Grad-CAM Visualizations}
Grad-CAM visualizations provided valuable insights into the key regions of chest X-rays that significantly influenced the model’s predictions\cite{Selvaraju19}. However, its sensitivity to noise in feature maps can result in inconsistent results. Future work could explore combining Grad-CAM with SHAP to enhance both visual and quantitative interpretability. Additionally, techniques like SmoothGrad \cite{Smilkov17} could further mitigate noise in feature maps, thereby improving the reliability of visualizations. Compared to other interpretability methods such as SHAP and LIME, Grad-CAM offers the unique advantage of directly visualizing class-specific activations in convolutional layers\cite{Lundberg17}, making it particularly well-suited for analyzing medical images. For example, Figure 5 highlights regions associated with atelectasis, aligning the model’s focus with clinically relevant areas. This enhanced interpretability bridges the gap between model predictions and medical expertise, reinforcing the model’s potential applicability in clinical workflows. Figures 5 and 6 highlight areas of clinical significance, such as regions indicative of atelectasis or effusion, aligning model attention with radiological expertise.  These visualizations allowed us to understand which parts of the image were most relevant to the model’s decision-making process, reinforcing the interpretability of the model in a medical context.

For example, Figure 5 shows the Grad-CAM visualization for positive Atelectasis using AlexNet, highlighting the areas that the model focused on.

\begin{figure}[ht]
\centering
\includegraphics[width=0.65\textwidth]{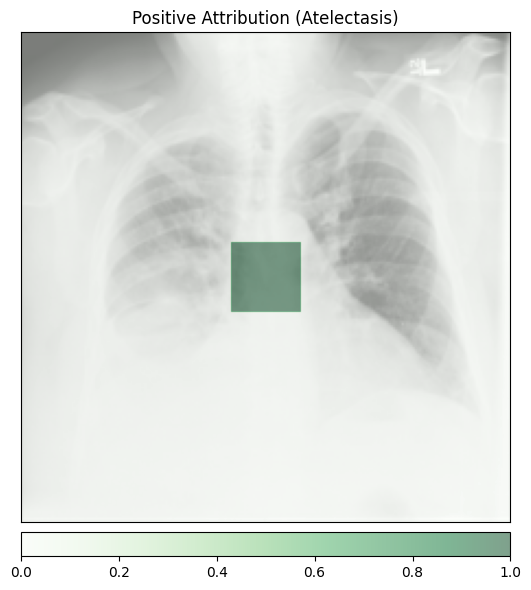} 
\caption{Grad-CAM for AlexNet - Positive Atelectasis}
\label{fig:gradcam_1}
\end{figure}

Similarly, Figure 6 presents the visualization for positive Effusion using AlexNet, demonstrating how the model identifies regions associated with fluid accumulation in the lungs.

\begin{figure}[ht]
\centering
\includegraphics[width=0.65\textwidth]{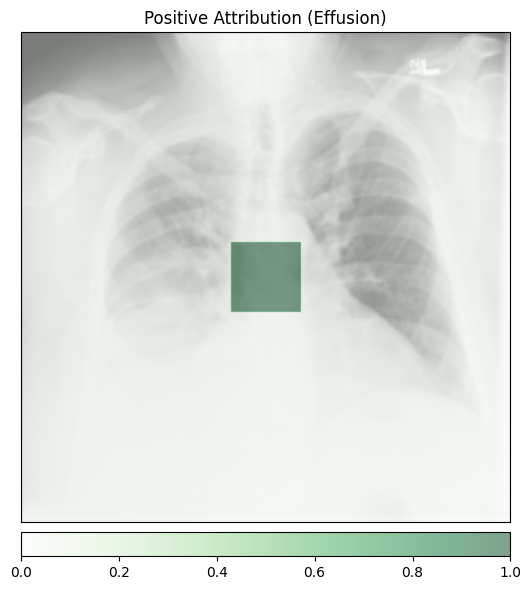} 
\caption{Grad-CAM for AlexNet - Positive Effusion}
\label{fig:gradcam_2}
\end{figure}

These visualizations confirm that the models are capable of attending to disease-relevant areas in the chest X-rays, which is crucial for model explainability in medical image classification.


\section{Conclusion} This study demonstrates the potential of deep learning in medical image classification, particularly for detecting thoracic diseases such as atelectasis and effusion. By leveraging transfer learning and Grad-CAM visualizations, the proposed approach achieves improved performance and interpretability, addressing key challenges in automated diagnostic systems. However, the models remain short of state-of-the-art performance, highlighting the need for further optimization.  By fine-tuning pre-trained models on a specialized medical dataset, we achieved promising results that highlight the feasibility of this approach. However, the models have not yet achieved state-of-the-art performance, underscoring the need for further optimization. Advancements in data preprocessing, the design of loss functions, and the adoption of more effective model architectures will be critical to enhancing generalization and ensuring clinical reliability.


\section{Future Work \& Challenges} Several areas require refinement to achieve optimal performance. Key improvements could be made by incorporating advanced data augmentation techniques, such as random rotation, flipping, and brightness adjustments, which would enhance model generalization and reduce overfitting. Additionally, exploring more advanced loss functions, specifically designed for multi-label classification, could improve how the model handles class imbalances. Future directions include exploring lightweight architectures like EfficientNet to reduce computational costs and experimenting with domain-specific pre-training to improve feature extraction\cite{Tan19}. Additionally, advanced data augmentation strategies, such as brightness and rotation adjustments, could mitigate overfitting and enhance model robustness. Finally, expanding the dataset to include diverse imaging modalities may further validate the clinical applicability of the proposed methods.

Despite these improvements, challenges such as class imbalance and overfitting persist. Although focal loss has been employed to address class imbalance, further strategies are needed to ensure underrepresented classes are better managed. Overfitting remains a concern, particularly when fine-tuning large pre-trained models on smaller datasets. Careful regularization will be critical to mitigating this risk. Additionally, while transfer learning can help alleviate training time, deep learning models still require substantial computational resources, which remains a practical challenge.

In the future, the research community can focus on optimizing data augmentation techniques, experimenting with hybrid loss functions, and exploring newer, more efficient model architectures to make significant progress in medical image classification. These advancements could lead to more robust, efficient models with greater potential for real-world clinical applications.

\section{Disclosures }
The authors declare no conflicts of interest.  

\section{Code, Data, and Materials Availability}
The software code and datasets supporting this study are available from the corresponding author upon reasonable request. 

\section{Acknowledgment}
All team members equally contributed to topic selection, data collection, coding, idea development, and the final report.



{\small
\bibliographystyle{ieee_fullname}
\bibliography{Manuscript}
}

\end{document}